\documentclass[onecolumn,showpacs,showkeys,amsmath,amssymb,superscriptaddress,floatfix,nofootinbib]{revtex4}

\usepackage{graphicx}
\usepackage{bm} 
\usepackage{subfigure}
\usepackage[colorlinks=true,linktocpage=true,linkcolor=blue,citecolor=blue]{hyperref}
\usepackage[usenames,dvipsnames]{color}

\makeatletter
\newcommand\erfc{\mathop{\operator@font erfc}\nolimits}
\def\slashchar#1{\setbox0=\hbox{$#1$}
   \dimen0=\wd0 \setbox1=\hbox{/} \dimen1=\wd1
   \ifdim\dimen0>\dimen1 \rlap{\hbox to \dimen0{\hfil/\hfil}} #1
   \else  \rlap{\hbox to \dimen1{\hfil$#1$\hfil}} / \fi}

\newcommand{\tildeN}{{\tilde N}}
\newcommand{\feq}{f_{\rm eq.}}
\newcommand{\taueq}{\tau_{\rm eq.}}

\newcommand{\mh}{{\hat m}}
\newcommand{\mheq}{\mh_{\rm eq.}}
\newcommand{\pres}{{\cal P}}
\newcommand{\preseq}{\pres_{\rm eq.}}
\newcommand{\ped}{{\cal E}}
\newcommand{\pedeq}{\ped_{\rm eq.}}
\newcommand{\Thetaeq}{\Theta_{\rm eq.}}
\newcommand{\Thetatr}{\Theta_{\sf tr.}}
\newcommand{\Thetatreq}{\Thetatr^{\rm eq.}}

\newcommand{\htr}{h_{\sf tr.}}
\newcommand{\hzero}{h_{\sf 0}}

\begin{document}
 
\title{
(3+1)-dimensional framework for leading-order non conformal anisotropic hydrodynamics 
}

\author{Leonardo Tinti} 
\email{tinti@fi.infn.it}
\affiliation{Institute of Physics, Jan Kochanowski University, PL-25406~Kielce, Poland}

\begin{abstract}

In this work I develop a new framework for anisotropic hydrodynamics that generalizes the leading order of the hydrodynamic expansion to the full (3+1)-dimensional anisotropic massive case. Following previous works, my considerations are based on the Boltzmann kinetic equation with the collisional term treated in the relaxation time approximation. The momentum anisotropy is included explicitly in the leading term, allowing for a large difference between the longitudinal and transverse pressures as well as for non trivial transverse dynamics.  Energy and momentum conservation is expressed by the first moment of the Boltzmann equation. The system of equations is closed by using the zeroth and second moments of the Boltzmann equation. The close-to-equilibrium matching with second-order viscous hydrodynamics is demonstrated. In particular,  I show that the coupling between shear and bulk pressure corrections, recently proved to be important for an accurate description of momentum anisotropy and bulk viscous dynamics, does not vanish in the close-to-equilibrium limit.

\end{abstract}

\pacs{12.38.Mh, 24.10.Nz, 25.75.-q, 51.10.+y, 52.27.Ny}

\keywords{relativistic heavy-ion collisions, quark-gluon plasma, anisotropic dynamics, viscous hydrodynamics, Boltzmann equation, relaxation time approximation, RHIC, LHC}

\maketitle 

\section{Introduction}
\label{sect:intro}

The successful application of relativistic viscous hydrodynamics in the description of heavy-ion collisions at RHIC (Relativistic Heavy-Ion Collider) and the LHC (Large Hadron Collider) has triggered a large interest in the development of the hydrodynamic framework
\cite{Israel:1976tn,Israel:1979wp,
Muronga:2001zk,Muronga:2003ta,
Heinz:2005bw,
Baier:2006um,Baier:2007ix,
York,Teanay,
Romatschke:2007mq,Dusling:2007gi,Luzum:2008cw,
Song:2008hj,El:2009vj,PeraltaRamos:2010je,
Denicol:2010tr,Denicol:2010xn,
Schenke:2010rr,Schenke:2011tv,
Bozek:2009dw,Bozek:2011wa,
Niemi:2011ix,Niemi:2012ry,
Bozek:2012qs,Denicol:2012cn,Jaiswal:2013npa,
PeraltaRamos:2012xk}. 
In this very active research field, several studies pointed out that viscous corrections combined with rapid longitudinal expansion (caused by large longitudinal gradients) result in a substantial pressure asymmetry at early times, even at a very low viscosity to entropy density ratio, $4\pi \eta/{\cal S} \sim 1$.  Similarly, a variety of microscopic models (string models, color glass condensate, pQCD kinetic calculations) predict large momentum anisotropies at early times. In fact, even in the limit of infinitely strong coupling, where the AdS/CFT correspondence can be used as an effective model, one finds a significant difference between the pressures in the transverse plane and in the longitudinal direction \cite{Heller:2011ju,Heller:2012je}. Interestingly, this difference slowly decays with time.  Large pressure anisotropies are a cause for concern, since viscous hydrodynamics treats pressure corrections in a perturbative manner and one may question the validity of the standard approximation scheme. Indeed, the presence of very large shear corrections (of the order of the isotropic pressure) may lead to unphysical results such as the negative longitudinal pressure and/or negative one-particle distribution functions. 

A new approach to treat these problems is {\it anisotropic hydrodynamics} (aHydro) \cite{Florkowski:2010cf,Martinez:2010sc,
Ryblewski:2010bs,Martinez:2010sd,
Ryblewski:2011aq,Martinez:2012tu,
Ryblewski:2012rr,Ryblewski:2013jsa,
Florkowski:2012ax,Florkowski:2012as,
Florkowski:2014txa,
Florkowski:2014sfa,
Florkowski:2011jg}. In this approach, in contrast with the conventional treatment, the effects connected with the expected high degree of pressure anisotropy of the produced matter are included in the leading order of the hydrodynamic expansion. In the original formulation of aHydro, the leading order includes a single anisotropy parameter taking into account a large difference between the longitudinal and transverse pressures. Unfortunately, this approach is unable to reproduce the pressure anisotropy in the transverse plane, which is generated by the radial flow \cite{Florkowski:2011jg}.  The transverse pressure anisotropy is important for the correct description of transverse collective behavior  and treatable, at least in the close to equilibrium limit, with the well-established framework of second-order viscous hydrodynamics. 

In Ref.~\cite{Bazow:2013ifa},  Bazow, Heinz, and Strickland extended the aHydro framework to the (2+1)-dimensional case (boost invariant but cylindrically asymmetric expansion, later denoted as the (2+1)D case). This formalism uses, however, the Romatschke-Strickland (RS) form~\cite{Romatschke:2003ms} of the distribution function at leading order, which implies that the transverse pressure asymmetries can be included only through  the second-order corrections. On the other hand, it has been demonstrated in Ref.~\cite{Tinti:2014conf} that it is possible to generalize the background, allowing nontrivial transverse dynamics already at leading order.  In Ref.~\cite{Tinti:2014conf} the leading order of anisotropic hydrodynamics for the (1+1)D case (boost invariant and cylindrically symmetric radial flow) was reformulated by introducing two independent anisotropy parameters, see also \cite{Florkowski:2014nc}. Afterwards, Nopoush, Ryblewski and Strickland introduced an additional parameter accounting for the isotropic pressure correction and better treatment of the bulk viscous pressure of massive systems~\cite{Nopoush:2014nc}.

In this paper, I present a new set of equations for anisotropic hydrodynamics in the (3+1)D case (a general three-dimensional expansion without constraints coming from boost invariance or cylindrical symmetry) including the effects of finite particle masses. Non-trivial transverse dynamics, i.e., a possibility that the two components of the transverse pressure are different, is naturally incorporated into the leading order of expansion. We show that the structure of our equations corresponds to that known from the recent second-order formulations of viscous hydrodynamics, if the system is close to local equilibrium. In particular, a coupling between shear and bulk viscous corrections is consistently reproduced. This coupling has been proven to be very important for a correct description of both pressure anisotropy and bulk viscous dynamics~\cite{Denicol:2014sbc, Jaiswal:2014}.

The paper is organized as follows: In the next Section, I discuss the Boltzmann equation in the relaxation time approximation, introduce the anisotropic phase-space distribution function, and discuss the zeroth and first moments of the kinetic equation. The anisotropic distribution function is characterized by a momentum scale parameter $\lambda$, a rank-two tensor $\xi^{\mu\nu}$ responsible for the pressure anisotropy, and a scalar parameter $\phi$ accounting for the isotropic pressure correction. The selection of the dynamic equations  is discussed in Sec.~\ref{sect:selection}. Section~\ref{sect:2mom} contains an analysis of the second moment of the Boltzmann equation. In Sec.~\ref{sect:close}  I analyze the close to equilibrium limit and discuss the matching with the second-order viscous hydrodynamics. I summarize and conclude in Sec.~\ref{sect:con}. Throughout the paper I am using the natural units, where $c=\hbar=k_B=1$ and the metric tensor has the signature $(+,-,-,-)$. Details of the calculations are presented in the Appendix.

\section{Boltzmann equation in the relaxation time approximation}
\label{sect:BE}

The basis of this work is the Boltzmann equation with the collisional kernel treated in relaxation-time approximation (RTA)~\cite{Bhatnagar:1954zz,Baym:1984np,Baym:1985tna,
Heiselberg:1995sh,Wong:1996va}
\begin{equation}
 p\cdot\partial f =  \frac{p\cdot U}{\taueq}
 \left( \frac{}{}\! \feq - f \right).
 \label{RTA}
\end{equation}
In Eq.~(\ref{RTA}), $f$ is the phase-space distribution function, $\tau_{\rm eq}$ is the relaxation time, and $f_{\rm eq}$ is the local equilibrium distribution function. For sake of simplicity, in the present work we consider a Boltzmann gas and neglect conserved charges. Thus, the local equilibrium distribution has the form

\begin{eqnarray}
f_{\rm eq.}(x,p) =  \tildeN\exp\left( - \frac{p \cdot U}{T} \right),
\label{feq}
\end{eqnarray}
where $T$ is the local equilibrium temperature and $\tildeN=N_{\rm dof}/(2\pi)^3$ is the normalization constant with $N_{\rm dof}$  being the number of internal degrees of freedom. The four-velocity $U^\mu$ in (\ref{feq}) is the eigenvector of the stress-energy tensor
\begin{equation}
 U_\mu T^{\mu\nu} = \ped U^\nu,
\end{equation}
which implies that we adopt the Landau prescription for $U$ while $\ped$ is the proper energy density.

\subsection{Anisotropic distribution function}
\label{sect:a-Hydro_ansatz}

\textcolor{black}{Anisotropic hydrodynamics is a reorganization of the hydrodynamics expansion around a non isotropic background. In order to obtain viscous hydrodynamics from an underlying kinetic theory, one usually expands the distribution function around the local equilibrium distribution}

\begin{equation}
 f = \feq +\delta f.
\end{equation}
\textcolor{black}{The deviation from local equilibrium $\delta f$ is assumed to be small, therefore justifying a perturbative treatment of the viscous corrections that depend upon it, namely the shear and bulk pressure corrections $\pi^{\mu\nu}$ and $\Pi$. The inverse Reynolds numbers, ${\sf R}^{-1}_{\Pi}= \Pi/\preseq$, ${\sf R}^{-1}_{\pi}=\sqrt{\pi_{\mu\nu}\pi^{\mu\nu}}/\preseq$, measure the validity of the perturbative treatment. In heavy ions collisions these numbers can be quite large, especially for the early dynamics.}

\textcolor{black}{In anisotropic hydrodynamics this problem is addressed by expanding the the distribution function around an anisotropic background }

\begin{equation}
 f = f_{\rm ansiso} +\delta {\tilde f}.
\end{equation}
\textcolor{black}{In this situation the pressure corrections read}

\begin{equation}\label{total_corrections}
 \pi^{\mu\nu} = \pi^{\mu\nu}_{\rm aniso} +{\tilde \pi}^{\mu\nu} \qquad \Pi = \Pi_{\rm aniso} + {\tilde \Pi}.
\end{equation}
\textcolor{black}{The anisotropic background is contributing to the pressure correction through $\pi^{\mu\nu}_{\rm aniso}$ and $\Pi_{\rm aniso}$, which are treated in a non perturbative approach. The modified Reynolds numbers ${\tilde {\sf R}}^{-1}_{{\tilde \Pi}}= {\tilde \Pi}/\preseq$, $\tilde{{\sf R}}^{-1}_{\tilde{\pi}}=\sqrt{{\tilde \pi}_{\mu\nu}\tilde{\pi}^{\mu\nu}}/\preseq$ have exactly the same role as the Reynolds numbers in viscous hydrodynamics, however it must be noted that $\tilde{\pi}^{\mu\nu}$ and $\tilde{\Pi}$ can be very small even when the total pressure corrections are large, as long as $\pi^{\mu\nu}_{\rm aniso}$ and $\Pi_{\rm aniso}$ can take into account the main contributions. In this way the anisotropic hydrodynamics expansion can still remain valid far away from local equilibrium, in contrast with the more familiar viscous hydrodynamics approach.}

\textcolor{black}{In this paper I will focus on the leading order of the anisotropic expansion $f_{\rm aniso}$, that is the zeroth order in the $\tilde{\pi}^{\mu\nu}$ and $\tilde{\Pi}$ corrections. From now on I will omit the $\cdots_{\rm aniso}$ part, understanding that the distribution function $f$ is the leading order $f_{\rm aniso}$. For more information about  the next to leading orders see Ref.~\cite{Bazow:2013ifa}.}

In the original formulations of aHydro, the basic assumption used is that the distribution function $f$ may be very well approximated by the RS form~\cite{Romatschke:2003ms}. The use of this form is, however, not satisfactory in the cases where the transverse expansion is taken into account and/or violations of boost invariance are included. In particular, viscous dynamics generates  different pressure corrections in different directions in the transverse plane, while the RS form allows for the difference between the longitudinal and transverse pressures only in the local rest frame. Having in mind these difficulties, in this work I will use a generalization of the RS form introduced in~\cite{Nopoush:2014nc}, namely
\begin{equation}
 f = \tildeN \exp\left( -\frac{1}{\lambda}\sqrt{ \frac{}{} p_\mu \Xi^{\mu\nu} p_\nu } \right),
\label{a-Hydro_distribution_0}
\end{equation}
where the anisotropy tensor $\Xi^{\mu\nu}$ reads
\begin{equation}
 \Xi^{\mu\nu} = U^\mu U^\nu + \xi^{\mu\nu} -\phi \Delta^{\mu\nu},
\label{decomposition}
\end{equation}
with $\xi^{\mu\nu}$ being the spatial, traceless part of $\Xi^{\mu\nu}$, and  $\phi$ being a degree of freedom controlling the trace of $\Xi^{\mu\nu}$. The projector $\Delta^{\mu\nu}$ is $ \Delta^{\mu\nu} \equiv g^{\mu\nu} - U^\mu U^\nu$.

In Eq.~(\ref{decomposition}) we neglect terms proportional to the products of space four vectors with $U$, since they are incompatible with the Landau prescription for the four-velocity. We do not multiply $U^\mu U^\nu$ by an extra scalar degree of freedom either, since this degree of freedom can be removed by rescaling other quantities, along with the momentum scale $\lambda$.

\textcolor{black}{This prescription is one of the simplest allowing for, possibly large, anisotropies and having the local equilibrium as a particular case. In the momentum space, the equal occupation number surface instead of being spherical, has an ellipsoidal form. Differently from previous prescriptions, in this work the direction having the largest deformation does not have to be exactly the longitudinal one.}

\subsection{Zeroth and first moments of Boltzmann equation}
\label{sect:01}

I am using a shorthand notation for the Lorentz invariant momentum measure
\begin{equation}
dP \equiv  \frac{d^3{\bf p}}{E} =   \frac{d^3{\bf p}}{\sqrt{m^2 + p^2}},
\label{Lorentz_invariant}
\end{equation}
with $m$ being the particle mass, and $p = (p_x^2 +p_y^2 + p_z^2)^{1/2}$. The zeroth moment of the kinetic equation (\ref{RTA}) is
\begin{eqnarray}
\int\!\! dP \; p^\mu\partial_\mu f =  \frac{1}{\taueq}U_\mu \int\!\! dP \,  p^\mu \left(\frac{}{} \!  \feq -f \right).
\label{zeroth_moment_0}
\end{eqnarray}
Because of the invariance of both the aHydro distribution function~(\ref{a-Hydro_distribution_0}) and the local equilibrium distribution~(\ref{feq}) with respect to the ${\bf p}\to-{\bf p}$ transformation in the local rest frame (LRF), we have
\begin{eqnarray}
N^\mu \equiv \int dP\, p^\mu f = n \,U^\mu, \quad 
N^\mu_{\rm eq.} = \int dP\, p^\mu \feq
= n_{\rm eq.}\, U^\mu.
\end{eqnarray} 
Therefore
\begin{eqnarray}
D n + n \theta = \frac{1}{\taueq} \left( \frac{}{} \!
n_{\rm eq.} - n \right),
\label{zeroth_moment_1}
\end{eqnarray}
where $D \equiv U^\mu \partial_\mu$ is the convective derivative (i.e., the time derivative along the fluid flux lines), and $\theta$ is the expansion scalar $\theta\equiv\partial_\mu U^\mu$. Having no particle conservation, however,  the particle density $n$ will be in general different from $n_{\rm eq.}$.

The first moment of the Boltzmann equation yields
\begin{equation}
 \int dP \, p^\mu p^\nu\,  \partial_\mu f = \frac{1}{\taueq} U_\mu\int dP \, p^\mu p^\nu \left( \frac{}{} \feq - f \right).
 \label{first_moment_0}
\end{equation}
Using the kinetic definition of the stress-energy-momentum tensor,  $T^{\mu\nu} = \int dP \, p^\mu p^\nu \, f$ (for instance, see~\cite{de_Groot}), the last equation can be rewritten as
\begin{equation}
 \partial_\mu T^{\mu\nu} = \frac{1}{\taueq}U_\mu \left( \frac{}{} T^{\mu\nu}_{\rm eq.} -T^{\mu\nu} \right) 
 = \frac{1}{\taueq}\left(\frac{}{} \ped_{\rm eq.} -\ped \right)U^\nu.
 \label{first_moment_1}
\end{equation}
The Landau matching condition,   $\ped=\ped_{\rm eq.}$, ensures local energy-momentum conservation and provides a formula that can be used to determine the effectve temperature $T$, namely
\begin{equation}
 \ped = \pedeq = 4\pi \tildeN \, T^4 \mheq^2 \left[ \frac{}{} 3K_2(\mheq)+ \mheq K_1(\mheq)\right],
\label{Landau_matching}
\end{equation}
where the notation $\mheq \equiv m/T$ has beeen used. The equilibrium particle density $n_{\rm eq.}$ and the equilibrium pressure $\preseq$ are then obtained from the expressions
\begin{eqnarray}\label{neq}
 && n_{\rm eq.} = 4\pi\tildeN\, T^3\mheq^2 K_2(\mheq), \\
 && \preseq = 4\pi\tildeN \, T^4\mheq^2 K_2(\mheq). 
\label{Peq}
\end{eqnarray}
Note that  $ \preseq$ (similarly to $n_{\rm eq.}$) is different from its non-equilibrium counterpart.

\section{Selection of dynamic equations}
\label{sect:selection}

There are eleven degrees of freedom in this formulation of anisotropic hydrodynamics: the effective temperature $T$, three independent components of the four-velocity $U^\mu$, the momentum scale $\lambda$, the scalar $\phi$, and five independent components of the space-like traceless symmetric rank two tensor $\xi^{\mu\nu}$.  In order to locally conserve energy and momentum, it is necessary to fulfill the equations coming from the first moment of the Boltzmann equation~(\ref{first_moment_1}) and the Landau matching condition~(\ref{Landau_matching}). The latter is an algebraic equation, completely defining the momentum scale $\lambda$ through the effective temperature $T$, the shear  anisotropy tensor $\xi^{\mu\nu}$, and the bulk parameter $\phi$. With the help of the Landau matching~(\ref{Landau_matching}), Eq.~(\ref{first_moment_1}) leads directly to local energy-momentum conservation
\begin{equation}
 \partial_\mu T^{\mu\nu} = 0.
\label{energy_momentum_conservation}
\end{equation}
One can use a general decomposition of $T^{\mu\nu}$ in the Landau frame
\begin{equation}
 T^{\mu\nu} = T_{\rm eq.}^{\mu\nu} + \pi^{\mu\nu} - \Pi \Delta^{\mu\nu} \quad = \quad \ped \, U^\mu U^\nu - \left( \frac{}{} \preseq +\Pi \right) \Delta^{\mu\nu} + \pi^{\mu\nu},
\label{Tmunu_decomposition}
\end{equation}
and obtain the time evolution of the proper energy density (connected directly with the effective temperature) by projecting (\ref{energy_momentum_conservation}) along the four-velocity $U_\nu$. In this way 
\begin{equation}
 D\pedeq = -\left( \pedeq +\preseq +\Pi \frac{}{} \right)\theta + \pi^{\mu\nu} \sigma_{\mu\nu},
\label{energy_conservation}
\end{equation}
where $\sigma^{\mu\nu} = \partial^{\langle\mu}U^{\nu\rangle}$. Here, the angular brackets are shorthand notation for the following tensor structure
\begin{equation}
 {\sf A}^{\langle\mu\nu\rangle} = \Delta^{\mu\nu}_{\alpha\beta} {\sf A}^{\alpha\beta} = \frac{1}{2}\left (\Delta^\mu_\alpha\Delta^\nu_\beta+\Delta^\nu_\alpha\Delta^\mu_\beta-\frac{2}{3}\Delta^{\mu\nu}\Delta_{\alpha\beta}\right){\sf A}^{\alpha\beta}.
\label{Deltamunualphabeta}
\end{equation}
On the other hand, the three independent components of the spatial part of Eq.~(\ref{energy_momentum_conservation}), i.e., the $\Delta^\mu_\nu$ projection, provide the time evolution of the four-velocity 
\begin{equation}
 \left( \pedeq + \preseq + \Pi \frac{}{}\right) DU^\mu = \nabla^\mu \preseq +\nabla^\mu \Pi +\Delta^\mu_\rho \partial_\sigma \pi^{\rho\sigma},
\label{momentum_conservation}
\end{equation}
where  $\nabla^\mu \equiv \Delta^{\mu\nu}\partial_\nu$ is the spatial gradient.

Accepting (\ref{energy_conservation}) and (\ref{momentum_conservation}) as a consequence of the energy-momentum conservation, there are still six extra equations needed for closing the system of dynamic equations, five for $\xi^{\mu\nu}$ and one for $\phi$. Unfortunately, there is no clear prescription for choosing these additional equations, as there are no other physical constraints that may directly fix all of them. Obviously, the zeroth moment is not sufficient to close the system of equations for $\xi^{\mu\nu}$ and $\phi$, and it is necessary to use higher moments. The second moment is a natural candidate as its tensor structure is sufficient to determine six independent equations.

In this work, I propose to use the zeroth and the second moment of the Boltzmann equation, and to choose the necessary equations by the requirement that they have the same geometric properties as corrections they have to describe. Hence, I first use the space-like, symmetric, and traceless part of the second moment.  The selection of the remaining scalar equation is less straightforward, so I am taking into account three possible cases: in the first case to take directly the zeroth moment, in the second case to use the trace of the space-like part of the second moment, finally, in the third case to choose the four-velocity projection of the second moment. All these cases have a close-to-equilibrium limit which agrees with second-order viscous hydrodynamics including the shear-bulk coupling (close to equilibrium, $\xi^{\mu\nu}$ and $\phi$ become proportional to the pressure corrections $\pi^{\mu\nu}$ and $\Pi$), as it will be discussed in Sec.~\ref{sect:close}. The difference remains in different forms of the (kinetic) coefficients which appear in the derived equations.  In this paper, the problem of the correct selection of the bulk equation will not be addressed further, and I leave it for further numerical study, where the predictions of this framework (with different options for the bulk dynamics) will be confronted with the exact solutions of the Boltzmann kinetic equations. This can be done in an analogous way as in Refs.~\cite{Florkowski:2013lza,Florkowski:2013lya}.

\section{Second moment}
\label{sect:2mom}

The second moment of the Boltzmann equation reads
\begin{equation}\label{II_mom_0}
 \partial_\lambda \Theta^{\lambda\mu\nu} = \frac{1}{\taueq} U_\lambda\left[ \frac{}{} \Thetaeq^{\lambda\mu\nu} - \Theta^{\lambda\mu\nu} \right].
\end{equation}
The rank three tensor $\Theta^{\lambda\mu\nu}$ is the third moment of the anisotropic distribution function $f$, while $\Thetaeq^{\lambda\mu\nu}$ is the third moment of the equilibrium distribution $f_{\rm eq.}$. Using Eq.~(\ref{a-Hydro_distribution_0}), because of ${\bf p}\to-{\bf p}$ invariance (LRF), it is possible to use the following decomposition
\begin{equation}\label{Theta}
 \Theta^{\lambda\mu\nu} = \int dP \, p^\lambda p^\mu p^\nu f = \Theta_U U^\lambda U^\mu U^\nu + U^\lambda \Theta^{\mu\nu} + U^\mu \Theta^{\nu\lambda} + U^\nu \Theta^{\lambda \mu},
\end{equation}
with the rank two tensor $\Theta^{\mu \nu}$ being
\begin{equation}
 \Theta^{\mu\nu} = U_\alpha \Delta^\mu_\beta \Delta^\nu_\gamma \, \Theta^{\alpha\beta\gamma}.
\end{equation}
The rank two tensor $ \Theta^{\mu\nu}$ itself can be written explicitly as a sum of the traceless and ``traceful'' part

\begin{equation}
 \Theta^{\mu\nu} = \Theta^{\langle\mu\nu\rangle} - \frac{1}{3}\Delta^{\mu\nu} \Theta_{\rm tr.},
\end{equation}
where
\begin{equation}
 \Thetatr= -\Delta_{\alpha\beta}\Theta^{\alpha\beta} = \int \!\! d^3{\bf p} \; p^2 \, f.
 \label{Thetatr}	
\end{equation}
Here, the sign convention has been chosen in such a way as to yield positive
 $\Thetatr$. The last equality in (\ref{Thetatr}) holds in the local rest frame.

\subsection{Shear equations}
\label{sect:shear}

In order that the number of equations obtained from the second moment is equal to the number of independent components in $\xi^{\mu\nu}$, I will take the traceless and orthogonal part of  the second moment with respect to the four-velocity $U$. In this way it reads
\begin{equation}
 \Delta^{\mu\nu}_{\alpha\beta}\partial_\lambda \Theta^{\lambda\alpha\beta} = \frac{1}{\taueq}U_\lambda\Delta^{\mu\nu}_{\alpha\beta}\left[ \frac{}{} \Thetaeq^{\lambda\alpha\beta} - \Theta^{\lambda\alpha\beta} \right] = -\frac{1}{\taueq}\Theta^{\langle\mu\nu\rangle},
\label{shear_0}
\end{equation}
where, in the last passage, it has been used the rotational invariance of the equilibrium distribution function in the local rest frame, resulting in a vanishing $\Thetaeq^{\langle\mu\nu\rangle}$. In the next step, using~(\ref{Theta}), the left-hand side of~(\ref{shear_0}) can be handled in the following way
\begin{equation}
  \Delta^{\mu\nu}_{\alpha\beta}\partial_\lambda \Theta^{\lambda\alpha\beta} =  \partial_\lambda\left( \frac{}{} \Delta^{\mu\nu}_{\alpha\beta} \Theta^{\lambda\alpha\beta} \right) - \Theta^{\lambda\alpha\beta}\partial_\lambda \Delta^{\mu\nu}_{\alpha\beta} = D\left( \frac{}{} \Delta^{\mu\nu}_{\alpha\beta}\Theta^{\alpha\beta} \right) + \theta \, \Theta^{\langle\mu\nu\rangle}- \Theta^{\lambda\alpha\beta}\partial_\lambda \Delta^{\mu\nu}_{\alpha\beta}.
\end{equation}
Using Eq.~(\ref{Theta}) again in the last expression, it is possible to rewrite Eq.~(\ref{shear_0}) as
\begin{equation}
 D\Theta^{\langle\mu\nu\rangle}+ \frac{5}{3}\theta \, \Theta^{\langle\mu\nu\rangle} +  2\Theta^{\langle\mu}_\lambda  \sigma^{\nu\rangle\lambda} -2\Theta^{\langle\mu}_\lambda \omega^{\nu\rangle\lambda} = -\frac{1}{\taueq} \Theta^{\langle\mu\nu\rangle},
\label{Shear_1}
\end{equation}
where I used the standard decomposition of the four velocity gradient
\begin{equation}
 \nabla_\mu U_\nu = \sigma_{\mu\nu} + \frac{1}{3}\theta\Delta_{\mu\nu} +\omega_{\mu\nu},
\end{equation}
with $\omega_{\mu\nu} \equiv( \nabla_{\mu}U_{\nu} - \nabla_{\nu}U_{\mu})/2 \equiv \nabla_{[\mu}U_{\nu]}$ being the vorticity.

\subsection{Bulk viscosity equations from the second moment}
\label{sect:bulk}

The tensor equation (\ref{Shear_1})  includes five independent equations needed to determine the shear dynamics. Therefore, it remains the the problem of finding the last missing equation which determines the bulk dynamics. The latter can be determined either from the zeroth moment or from a combination of the equations included in the second moment.

It is possible to construct two scalar equations from the second moment in a natural way by contracting it with $\Delta_{\mu\nu}$ or with $U_\mu U_\nu$.

Considering now the first possibility for selecting the necessary equation, namely the trace of Eq.~(\ref{II_mom_0})
\begin{equation}\label{bulk_1}
 \Delta_{\mu\nu} \partial_\lambda\Theta^{\lambda\mu\nu} = -\frac{1}{\taueq}\left[ \frac{}{}  \Thetatr^{\rm eq.} - \Thetatr \right].
\end{equation}
Following the same procedure as was used for the shear equations, the last equation reads
\begin{equation}
 \Delta_{\mu\nu} \partial_\lambda\Theta^{\lambda\mu\nu} =  -\partial_\lambda\left( \frac{}{} U^\lambda \Thetatr \right) - \Theta^{\lambda\mu\nu}\partial_\lambda \Delta_{\mu\nu} = -D\Thetatr - \theta \, \Thetatr -\Theta^{\lambda\mu\nu}\partial_\lambda \Delta_{\mu\nu},
\end{equation}
which finally leads to
\begin{equation}\label{bulk_2}
D\Thetatr + \frac{5}{3} \, \theta \, \Thetatr - 2 \Theta_{\langle\mu\nu\rangle}\sigma^{\mu\nu} = \frac{1}{\taueq}\left[ \frac{}{} \Thetatreq - \Thetatr  \right].
\end{equation}
Alternatively,  the $U_\mu U_\nu$ projection of the second moment gives
\begin{equation}\label{UmuUnu_0}
 U_\mu U_\nu \partial_\lambda\Theta^{\lambda\mu\nu} = \frac{1}{\taueq}\left[ \frac{}{} \Theta_U^{\rm eq.} - \Theta_U\right].
\end{equation}
Using the definition of $\Theta_U$ in~(\ref{Theta}), it is possible to write
\begin{equation}\label{ThetaU}
 \Theta_U = \int dP \, (p^\mu U_\mu)^3 \, f = \int dP \, (p^\mu U_\mu)\left[ m^2 - p^\mu \Delta_{\mu\nu}  p^\nu \right] = m^2 \, n +\Thetatr.
\end{equation}
A very similar relation holds for the left-hand side of Eq.~(\ref{UmuUnu_0}), namely
\begin{eqnarray}\nonumber
 U_\mu U_\nu \partial_\lambda\Theta^{\lambda\mu\nu} &=& \partial_\lambda \left( \frac{}{} U^\lambda \Theta_U \right) -\Theta^{\lambda\mu\nu}\partial_\lambda\left( \frac{}{}  U_\mu U_\nu \right) \\ \nonumber
 &=& D\Theta_U +\theta\Theta_U + \Theta^{\lambda\mu\nu}\partial_\lambda\Delta_{\mu\nu}  \\ \nonumber
 &=& m^2\left( Dn + n \, \theta \right) + \left[ D\Thetatr + \theta \, \Thetatr +\Theta^{\lambda\mu\nu}\partial_\lambda \Delta_{\mu\nu}\right]  \\ 
 &=& m^2\left( Dn + n \, \theta \right) + \left[ D\Thetatr + \frac{5}{3} \, \theta \, \Thetatr - 2 \Theta_{\langle\mu\nu\rangle}\sigma^{\mu\nu} \right].
\end{eqnarray}
Therefore, Eq.~(\ref{UmuUnu_0}) is a linear combination of the zeroth moment of the Boltzmann equation and Eq.~(\ref{bulk_1})
\begin{eqnarray}\label{UmuUnu_1}
  m^2\left( \frac{}{}Dn + n \, \theta \right) +\left[ D\Thetatr + \frac{5}{3} \, \theta \, \Thetatr - 2 \Theta_{\langle\mu\nu\rangle}\sigma^{\mu\nu}\right] = \frac{1}{\taueq}\left\{ m^2\left[ \frac{}{} n_{\rm eq.} -n \right] + \left[ \frac{}{} \Thetatreq -\Thetatr \right] \right\}.
\end{eqnarray}
In general, any linear combination of Eq.~(\ref{zeroth_moment_1}) and~(\ref{bulk_2}) is acceptable, not only Eq.~(\ref{UmuUnu_1}). In the next section I will show that all three considered cases have the correct close-to-equilibrium limit, leaving the selection of the best equation for further study.

\section{Close-to-equilibrium behavior and pressure corrections}
\label{sect:close}

In the close-to-equilibrium limit, the anisotropy tensor $\xi^{\mu\nu}$ and the scalar $\phi$ defined by the formula~(\ref{decomposition}) become proportional to the shear and bulk pressure corrections, $\pi^{\mu\nu}$ and $\Pi$, respectively. The latter are defined by the standard decomposition of the stress energy momentum tensor with the Landau prescription for the four velocity in~(\ref{Tmunu_decomposition}). The connection between $\pi^{\mu\nu}$ and $\xi^{\mu\nu}$ can be easily seen using geometrical arguments, just considering the linear contribution in the allegedly small anisotropy parameters. On the other hand, since there are two scalar quantities for the non-equilibrium distribution, the proportionality between $\Pi$ and $\phi$ is less trivial. However, expanding the anisotropic distribution~(\ref{a-Hydro_distribution_0}) around the equilibrium one,
\begin{equation}\label{a-Hydro_distribution->eq}
 f \simeq \feq \left[ 1 + \frac{\lambda -T}{T^2}(p\cdot U) -\frac{1}{2T(p\cdot U)}\left( \frac{}{} \xi_{\mu\nu} - \phi \Delta_{\mu\nu} \right) p^\mu p^\nu \right],
\end{equation}
and make use of the Landau matching~(\ref{Landau_matching}), it is possible to remove the $\lambda - T$ correction and prove the following expansion
\begin{equation}
 T^{\mu\nu} \simeq T^{\mu\nu}_{\rm eq.} - \varpi_{\xi}(m,T)\xi^{\mu\nu} + \varPi_\phi(m,T)\phi. 
\label{T->eq}
\end{equation}
\textcolor{black}{Therefore, up to the leading order, the pressure corrections read}
\begin{eqnarray}
 \Pi = -\varPi_\phi(m,T)\phi, \qquad   \pi^{\mu\nu}= -\varpi_\xi(m,T)\xi^{\mu\nu}.
\label{Pp}
\end{eqnarray}
For the definition of the functions $\varpi_\xi$ and $\varPi_\phi$ in (\ref{Pp}) see Appendix~\ref{sect:cte}.

Equations~(\ref{Shear_1}) look quite  similar to the second-order viscous hydrodynamics equations for the shear pressure correction. Equation~(\ref{zeroth_moment_1}) is the particle creation equation in  RTA, while Eq.~(\ref{bulk_2}) and, to a lesser extent, Eq.~(\ref{UmuUnu_1}) resemble equations for the isotropic pressure correction $\Pi$.

 \textcolor{black}{Indeed, as it will be shown below, Eq~(\ref{Shear_1})} reduces to  

\begin{equation}
 \tau_\pi D\pi^{\langle\mu\nu\rangle} + \pi^{\mu\nu} = 2\eta \, \sigma^{\mu\nu} + 2 \, \tau_\pi \, \pi_\lambda^{\langle\mu}\omega^{\nu\rangle\lambda} -\tau_{\pi\pi}  \: \pi_\lambda^{\langle\mu}\sigma^{\nu\rangle\lambda} - \delta_{\pi\pi} \, \pi^{\mu\nu}\theta + \lambda_{\pi\Pi} \, \Pi \, \sigma^{\mu\nu} + \phi_6 \, \Pi\,\pi^{\mu\nu}  + \phi_7 \, \pi_\lambda^{\langle\mu}\pi^{\nu\rangle\lambda},
\label{visc_shear}
\end{equation}
%
\textcolor{black}{in the close-to-equilibrium limit, and all the scalar equations~(\ref{zeroth_moment_1}),~(\ref{bulk_2}) and~(\ref{UmuUnu_1}) reduce to}

\begin{equation}
 \tau_\Pi \, D\Pi +  \Pi =  -\zeta \, \theta  - \delta_{\Pi\Pi} \, \Pi \, \theta + \lambda_{\Pi\pi} \, \pi^{\mu\nu}\sigma_{\mu\nu} +\phi_1\, \Pi^2 +\phi_3 \, \pi_{\mu\nu}\pi^{\mu\nu}.
\label{visc_bulk}
\end{equation}
\textcolor{black}{Both Eq.~(\ref{visc_shear}) and~(\ref{visc_bulk}) are a particular case of the most general equation for shear and bulk viscosity that can be extracted from kinetic theory using the approach of Ref.~\cite{Denicol:2012cn}, up to the second order in the Knudsen\footnote{The Knudsen number is the ratio between the microscopic and macroscopic scale $l_{\rm micro.}/L_{\rm macro.}$. It is usually thought to be the same for both time and space scales, and the derivatives of macroscopic quantities are supposed to be proportional to it. Therefore the order of gradients of a term equals the power in the Knudsen number.} and/or inverse Reynolds numbers. It is interesting to note that they have a similar structure of the very successful prescriptions for viscous hydrodynamics, which have been tested in Ref.~\cite{Jaiswal:2014}, with the difference of having non-vanishing $\phi_1$, $\phi_3$, $\phi_6$ and $\phi_7$ transport coefficients. They all include the bulk-shear couplings, necessary for the correct description of the isotropic pressure correction (see Ref.~\cite{Denicol:2014sbc}).}

Making use of the expansion~(\ref{a-Hydro_distribution->eq}), in the close-to-equilibrium limit one may write
\begin{equation}
 \Theta^{\langle\mu\nu\rangle}\simeq g(m,T)\pi^{\mu\nu}, \qquad \Thetatr \simeq \Thetatreq +\htr(m,T)\,  \Pi, \qquad n \simeq n_{\rm eq.} - \hzero(m,T) \, \Pi.
\label{q->eq} 
\end{equation}
Consequently, the derivatives of these expressions are as follows
\begin{eqnarray}\nonumber
 D \Theta^{\langle\mu\nu\rangle} &=& \Delta^{\mu\nu}_{\alpha\beta}D\Theta^{\alpha\beta} \simeq  \Delta^{\mu\nu}_{\alpha\beta}D\left[ g\, \pi^{\alpha\beta} -\frac{1}{3}\left( \frac{}{} \Thetatreq +\htr\,\Pi \right)\Delta^{\alpha\beta} \right] = g \, \Delta^\mu_\alpha\Delta^\nu_\beta \, D\pi^{\alpha\beta} + \pi^{\mu\nu} \, Dg, \\ \label{derivatives->eq}
 D\Thetatr &\simeq& D\left( \frac{}{} \Thetatreq + \htr \, \Pi \right)=  \htr \, D\Pi + \Pi\, D\htr + D\Thetatreq, \\ \nonumber
 Dn &\simeq& -\hzero D\Pi - \Pi \, D\hzero + Dn_{\rm eq.},
\end{eqnarray}
\textcolor{black}{In order to have a meaningful comparison with second order viscous hydrodynamics it is necessary to include a second order expansion of $\Theta^{\mu\nu}$ and  $n$. On the left hand side of Eqs.~(\ref{zeroth_moment_1}),~(\ref{Shear_1}),~(\ref{bulk_2}) and~(\ref{UmuUnu_1}), all of the terms are multiplied by gradients of the four velocity, or they appear as a convective derivative. They are therefore at least of first order in gradients ({\it i.e.} first order in Knudsen number). On the other hand the right hand side is zeroth order in Knudsen number. Neglecting the second order corrections, some second order terms on pressure corrections would be systematically neglected.  In the physically meaningful case where both the pressure corrections and the gradients are small, these corrections are expected to be of the same order of magnitude as the other  considered. Taking into account the next to leading order in the close-to-equilibrium limit, Eqs.~(\ref{q->eq}) read}

\begin{eqnarray}\label{q->eq_II} 
 \Theta^{\langle\mu\nu\rangle}&\simeq& g(m,T)\left\{ \frac{}{} \pi^{\mu\nu} - \phi_6 \; \Pi \, \pi^{\mu\nu} - \phi_7 \; \pi^{\langle\mu}_\lambda \pi^{\nu\rangle\lambda}\right\}, \\ \nonumber
 \Thetatr &\simeq& \Thetatreq +\htr(m,T)\left\{ \frac{}{} \Pi -\phi_1^{\sf tr.} \, \Pi^2 -\phi_3^{\sf tr.}\, \pi_{\mu\nu}\pi^{\mu\nu} \right\}, \\ \nonumber
 n &\simeq& n_{\rm eq.} - \hzero(m,T) \left\{ \Pi -\phi_1^{0} \; \Pi^2 -\phi_3^{0}\; \pi_{\mu\nu}\pi^{\mu\nu} \right\},
\end{eqnarray}
\textcolor{black}{for the definitions of $g$, $\htr$, $\hzero$, $\phi_6$, $\phi_7$, $\phi_1^{\sf tr.}$, $\phi_3^{\sf tr.}$, $\phi_1^{0}$ and $\phi_3^{0}$ see Appendix~\ref{sect:cte}.}

\textcolor{black}{Using Eqs.~(\ref{q->eq_II}) and~(\ref{derivatives->eq}), the shear equation~(\ref{Shear_1}) reads}
\begin{eqnarray}\label{shear->eq_1}
 D\pi^{\langle\mu\nu\rangle} +\frac{1}{\taueq}\pi^{\mu\nu} &=& \frac{2}{3}\frac{\Thetatreq}{g}\sigma^{\mu\nu} + 2 \, \pi^{\langle\mu}_\lambda \omega^{\nu\rangle\lambda} -\frac{5}{3} \, \theta \, \pi^{\mu\nu} - 2 \, \pi^{\langle\mu}_\lambda \sigma^{\nu\rangle\lambda} + \frac{2}{3} \frac{\htr}{g} \, \Pi \, \sigma^{\mu\nu} -  D\ln (g) \, \pi^{\mu\nu} \\ \nonumber
 && \qquad +\frac{\phi_6}{\taueq} \Pi \, \pi^{\mu\nu} + \frac{\phi_7}{\taueq} \pi^{\langle\mu}_\lambda \pi^{\nu\rangle\lambda}.
\end{eqnarray}
The term $D\ln(g)$ contains derivatives of the temperature $T$. In order to remove them, it can be used the exact relation
\begin{equation}
 DT = -\frac{\pedeq +\preseq}{\partial_T\pedeq}\, \theta \; - \; \frac{1}{\partial_T\pedeq}\, \theta \, \Pi \;+ \;\frac{1}{\partial_T\pedeq} \sigma_{\mu\nu}\pi^{\mu\nu},
\label{DT}
\end{equation}
which stems directly from Eq.~(\ref{energy_conservation}), plugging the formula for the energy density in~(\ref{Landau_matching}). Therefore, the convective derivative $D\ln(g)$ reads
\begin{equation}
 D\ln(g) = \frac{\partial_T g}{g}  DT =  -\frac{\partial_T g}{g}\left( \frac{\pedeq +\preseq}{\partial_T\pedeq}\right) \theta \; - \; \frac{\partial_T g}{g \, \partial_T\pedeq}\, \theta \, \Pi \;+ \;\frac{\partial_T g}{g \, \partial_T\pedeq} \sigma_{\mu\nu}\pi^{\mu\nu}.
\end{equation}
Keeping only the linear correction in $\xi^{\mu\nu}$ and $\phi$ ($\pi^{\mu\nu}$ and $\Pi$), Eq.~(\ref{shear->eq_1}) can be rewritten as
\begin{eqnarray}\label{shear->eq_2}
 D\pi^{\langle\mu\nu\rangle} +\frac{1}{\taueq}\pi^{\mu\nu} &=& \frac{2}{3}\frac{\Thetatreq}{g}\sigma^{\mu\nu} + 2 \, \pi^{\langle\mu}_\lambda \omega^{\nu\rangle\lambda} -\left[ \frac{5}{3} - \frac{\partial_T g}{g}\left( \frac{\pedeq +\preseq}{\partial_T\pedeq}\right) \right] \theta \, \pi^{\mu\nu} - 2 \, \pi^{\langle\mu}_\lambda \sigma^{\nu\rangle\lambda} + \frac{2}{3} \frac{\htr}{g} \, \Pi \, \sigma^{\mu\nu} \\ \nonumber 
 && \qquad +\frac{\phi_6}{\taueq} \Pi \, \pi^{\mu\nu} + \frac{\phi_7}{\taueq} \pi^{\langle\mu}_\lambda \pi^{\nu\rangle\lambda}.
\end{eqnarray}
The situation is very similar for the scalar equations. Indeed, \textcolor{black}{up to second order}, Eq.~(\ref{zeroth_moment_1}) reads 

\begin{equation}
 D\Pi + \frac{1}{\taueq} \, \Pi = \frac{n_{\rm eq.}}{\hzero} \, \theta -\theta \, \Pi + \frac{1}{\hzero}D n_{\rm eq.} - D\ln(\hzero) \, \Pi + \frac{\phi_1^0}{\taueq} \Pi^2 + \frac{\phi_3^0}{\taueq} \pi_{\mu\nu}\pi^{\mu\nu}.
\end{equation}
Both $\hzero$ and $n_{\rm eq.}$ are functions of the temperature $T$. For calculating their convective derivative it is possible to use again Eq.~(\ref{DT}) and neglect the terms which are non-linear in the pressure corrections. Hence the particle creation equation~(\ref{zeroth_moment_1}) reads
\begin{eqnarray}\label{zeroth->eq}
 D\Pi + \frac{1}{\taueq} \, \Pi &=& -\frac{1}{\hzero}\left( \partial_T n_{\rm eq.} \frac{\pedeq + \preseq}{\partial_T \pedeq} - n_{\rm eq.}\right) \, \theta -\left( 1 + \frac{\partial_T n_{\rm eq.} }{\hzero \, \partial_T \pedeq} - \frac{\partial_T \hzero}{\hzero} \, \frac{\pedeq + \preseq}{\partial_T \pedeq} \right)\theta \, \Pi \\ \nonumber
 && \quad  + \frac{\partial_T n_{\rm eq.} }{\hzero \, \partial_T \pedeq } \, \pi_{\mu\nu}\sigma^{\mu\nu} + \frac{\phi_1^0}{\taueq} \Pi^2 + \frac{\phi_3^0}{\taueq} \pi_{\mu\nu}\pi^{\mu\nu},
\end{eqnarray}
the formula~(\ref{bulk_2}) becomes
\begin{eqnarray}\label{bulk->eq}
 D\Pi +\frac{1}{\taueq}\Pi &=&  -\frac{1}{\htr}\left( \frac{5}{3}\Thetatreq - \partial_T \Thetatreq \,  \frac{\pedeq + \preseq}{\partial_T \pedeq} \right) \, \theta - \left( \frac{5}{3} - \frac{\partial_T\Thetatreq}{\htr \, \partial_T\pedeq} - \frac{\partial_T \htr}{\htr} \, \frac{\pedeq + \preseq}{\partial_T\pedeq}\right) \theta \, \Pi  \\ \nonumber
 && \qquad +\frac{1}{\htr}\left( 2g - \frac{\partial_T\Thetatreq}{\partial_T\pedeq} \right) \pi_{\mu\nu}\sigma^{\mu\nu} + \frac{\phi_1^{\sf tr.}}{\taueq} \Pi^2 + \frac{\phi_3^{\sf tr.}}{\taueq} \pi_{\mu\nu}\pi^{\mu\nu},
\end{eqnarray}
while Eq.~(\ref{UmuUnu_1}) has the form
\begin{eqnarray}\nonumber
 && D\Pi + \frac{1}{\taueq}\Pi = -\frac{1}{\htr -m^2\hzero}\left[  \frac{5}{3} \Thetatreq + m^2 n_{\rm eq.} - \left(  \frac{}{}  \partial_T \Thetatreq + m^2 \partial_Tn_{\rm eq.}  \right)\frac{\pedeq + \preseq}{\partial_T \pedeq}  \right] \, \theta  \\ \nonumber
 && - \frac{1}{\htr -m^2\hzero}\left\{ \frac{}{} \frac{5}{3} \htr - m^2 \hzero - \left[   \partial_T \Thetatreq + m^2 \partial_T n_{\rm eq.} + \left(  \frac{}{} \partial_T \htr-m^2\partial_T \hzero \right)\left( \frac{}{} \pedeq + \preseq \right)    \right]\frac{1}{\partial_T \pedeq}  \right\} \, \theta \, \Pi  \\ \nonumber
  && + \frac{1}{\htr -m^2\hzero} \left[ 2 \, g - \left(  \frac{}{} \partial_T \Thetatreq + m^2 \partial_Tn_{\rm eq.}  \right)\frac{1}{\partial_T \pedeq} \right]\, \pi_{\mu\nu}\sigma^{\mu\nu} \\
  && + \frac{\htr \, \phi_1^{\sf tr.} -m^2 \, \hzero \, \phi_1^0}{\taueq\left( \htr-m^2 \,\hzero \right)} \; \Pi^2 + \frac{\htr \, \phi_3^{\sf tr.} -m^2 \, \hzero \, \phi_3^0}{\taueq\left( \htr-m^2 \, \hzero \right)} \; \pi_{\mu\nu}\pi^{\mu\nu}. \label{UmuUnu->eq}
\end{eqnarray}
It is important to stress that for systems close to equilibrium the structure of the proposed equations 
\textcolor{black}{is similar} to the one used in the very successful second-order hydrodynamics prescriptions discussed in Ref.~\cite{Jaiswal:2014}. 
\textcolor{black}{They all include }the shear to bulk and bulk to shear couplings, absent in the standard Israel Stewart approach but necessary for the correct description of the $(0+1)$D Bjorken dissipative expansion~\cite{Denicol:2014sbc}. It is also interesting to note that the second-order viscous hydrodynamics formulations discussed in Refs.~\cite{Denicol:2014sbc, Jaiswal:2014} and Eqs.~(\ref{shear->eq_1}),~(\ref{zeroth->eq}),~(\ref{bulk->eq}) and~(\ref{UmuUnu->eq}) in the present work, all provide the same shear to vorticity coupling, and the same Navier-Stokes limit (first order viscous hydrodynamics); i.e. the values for the shear viscosity and bulk viscosity are numerically the same, despite the very different  manner in which they have been extracted. \textcolor{black}{Another interesting comparison is the massless limit, where Eq.~(\ref{shear->eq_2}) reads}

\begin{eqnarray}\label{shear->eq_conf}
 D\pi^{\langle\mu\nu\rangle} +\frac{1}{\taueq}\pi^{\mu\nu} &=& 2  \frac{\eta}{\taueq} \sigma^{\mu\nu} + 2 \, \pi^{\langle\mu}_\lambda \omega^{\nu\rangle\lambda} -\frac{4}{3}\theta \, \pi^{\mu\nu} - 2 \, \pi^{\langle\mu}_\lambda \sigma^{\nu\rangle\lambda} + \frac{2}{7}\frac{1}{2\eta} \pi^{\langle\mu}_\lambda \pi^{\nu\rangle\lambda}.
\end{eqnarray}
\textcolor{black}{In the conformal limit, both prescriptions for second order viscous hydrodynamics tested in Refs.~\cite{Denicol:2014sbc, Jaiswal:2014} give the same transport coefficients, which agree with the results given in Ref.~\cite{Teanay} as long as the system is close to the Navier-Stokes solutions and it is possible to use the approximation $\sigma^{\mu\nu}\simeq\pi^{\mu\nu}/2\eta$ in the shear to shear correction. There is agreement between these transport coefficients and the ones in Eq.~(\ref{shear->eq_conf}), except for the last two, which can be grouped together in a single one in this limit. The shear to shear correction is then $6/5$ larger then the one in Ref.~\cite{Teanay}. This is not completely unexpected. The pressure corrections~(\ref{total_corrections}) are not fully described by the leading order in the anisotropic expansion. The consequences of this disagreement are not clear at the moment, however it is reasonable to expect that the non non linear treatment of the pressure corrections in anisotropic hydrodynamics can compensate for this shortcoming. The previous version of anisotropic hydrodynamics  (see Refs.~\cite{Tinti:2014conf, Nopoush:2014nc, Nopoush:2015gf}), is not matching with second order viscous hydrodynamics in (3+1)D, however, there is a matching in (1+1)D. If we perform a second order expansion in the anisotropy parameters we recover the same, larger, shear to shear correction. Nonetheless there is a striking agreement  with the exact solution of the Boltzmann equation, both for the Bjorken flow~\cite{Florkowski:2014nc,Nopoush:2014nc,Denicol:2014sbc}, and for the Gubser flow~\cite{Nopoush:2015gf}.}  

\section{Summary and conclusions}
\label{sect:con}

In this paper, I have introduced a new formulation of anisotropic hydrodynamics for massive fluids. In contrast to previous works, I have not used any simplifying assumptions, such as the longitudinal boost invariance or cylindrical symmetry for the transverse expansion, providing a full $(3+1)$D framework at leading order. On the other hand, my considerations have been based again on the Boltzmann kinetic equation with the collision term treated in the relaxation-time approximation.

The newly developed framework of anisotropic hydrodynamics includes eleven equations --- one for each degree of freedom from the set containing: the temperature $T$, the three independent components of the four-velocity $U^\mu$, the momentum scale $\lambda$, the bulk degree of freedom $\phi$, and five independent components of the spacelike, traceless, and symmetric tensor $\xi^{\mu\nu}$. In order to locally conserve four-momentum, it is necessary to include among the proposed equations the first moment of the Boltzmann equation~(\ref{first_moment_1}) and the Landau matching~(\ref{Landau_matching}). There are no obvious other physical prescriptions to choose the remaining six equations. In this work, I have considered only the zeroth and the second moment of the Boltzmann equation as a source for the extra equations; having in mind that the second moment is the first in the sequence of moments which has enough independent components to close the system. As in our previous work~\cite{Tinti:2014conf} I have proposed to use the second moment for describing the shear pressure correction, and I have proved that Eqs.~(\ref{Shear_1}), when considered in the close-to-equilibrium limit, take a familiar form of the second-order viscous hydrodynamics equations for the shear pressure corrections.

Using a simple argument of matching between the geometrical properties of the proposed equation and the correction this equation should describe, I have found that there is more than one option for the scalar equation describing the bulk viscous dynamics. Consequently,  in this work I have examined three possible cases as possible equations for the isotropic pressure correction:  i)  the particle creation equation~(\ref{zeroth_moment_1}), i.e.,  the zeroth moment, ii) the trace of the second moment of the Boltzmann equation~(\ref{bulk_2}) and, finally, iii)  the four velocity projection~(\ref{UmuUnu_1}) of the second moment. All of them provide the correct close to equilibrium limit, see Eqs.~(\ref{zeroth->eq}),~(\ref{bulk->eq}), and~(\ref{UmuUnu->eq}), respectively.  An interesting way to select one of the three formulations would be to confront their results with the exact solutions of the Boltzmann kinetic equation, in an analogous way as  it has been done before in Refs.~\cite{Florkowski:2013lza,Florkowski:2013lya}.

\acknowledgments

I thank Wojciech Florkowski, Rados\l aw Ryblewski and Michael Strickland for clarifying discussions and interesting comments.  This work has been supported by Polish National Science Center grant No. DEC-2012/06/A/ST2/00390.

\section{Appendix: Explicit formulas}
\label{sect:explicit}

\subsection{Anisotropic expansion}
\label{sect:explicit}

In this Appendix I collect the exact expressions for the integrals used in the anisotropic hydrodynamics equations. In order to do so, it is convenient to use a smooth orthonormal basis $ \{U,X_i\}$, with the four-velocity $U$ as the time direction, and three vector fields fulfilling
\begin{equation}
 U_\mu X_i^\mu = 0, \quad \forall i \qquad\qquad g_{\mu\nu} \, X_i^\mu X_j^\nu = -\delta_{ij}.
\end{equation}
For instance, it is possible to take the lab frame spatial directions and obtain the $X_i^\mu$ four vectors using the Gram-Schmidt orthonormalization procedure. The stress-energy-momentum tensor then reads
\begin{equation}
 T^{\mu\nu} = \ped U^\mu U^\nu + \pres^{ij}X^\mu_i X^\nu_j.
\end{equation}
Starting with the proper energy density, in the local rest frame it reads
\begin{equation}
 \ped = \int dP \; (p\cdot U)^2 \, f = \tildeN \int d^3{\bf p} \; \sqrt{  m^2 + p^2} \, \exp\left\{ -\frac{1}{\lambda}\sqrt{ \frac{}{} m^2 + {\bf p}\cdot M\cdot {\bf p}} \right\},
\end{equation}
where the $3\times 3$ matrix $ M$ is
\begin{equation}
 M_{ij} = \xi_{ij} + (1+\phi)\delta_{ij}.
\end{equation}
It is important to note that $M$ must be positive-definite, otherwise there would be an imaginary part of the Boltzmann function. That means, that there exists a rotation ${\sf R}$ diagonalizing the matrix $M$,
\begin{equation}
 {\sf R}\cdot M\cdot {\sf R}^{-1} = {\rm diag}(\alpha_1,\alpha_2,\alpha_3),
\end{equation}
where the eigenvalues $\alpha_i$ are all positive. The energy density thus reads

\begin{eqnarray}\nonumber
 \ped &=& \tildeN \int d^3{\bf p} \; \sqrt{  m^2 + p^2} \, \exp\left\{ -\frac{1}{\lambda}\sqrt{ \frac{}{} m^2 + \sum_k \alpha_k (p^k)^2} \right\} = \\
 &=& \frac{\tildeN}{\sqrt{\alpha_1\alpha_2\alpha_3}} \int d^3{\bf p} \; \sqrt{  m^2 + \sum_k\frac{(p^k)^2}{\alpha_k}} \, \exp\left\{ -\frac{1}{\lambda}\sqrt{ m^2 + p^2 } \right\}.
\end{eqnarray}
Performing another rotation back to the original frame, one finds
\begin{equation}
 \ped = \sqrt{ \det(M^{- 1})} \int d^3{\bf p} \; \sqrt{\frac{}{}  m^2 + {\bf p}\cdot (M^{-1})\cdot{\bf p} } \; \feq(\lambda),
\end{equation}
where  $\feq(\lambda)$ is the local equilibrium distribution with the momentum scale $\lambda$ instead of the temperature $T$. It is straightforward to check that this formula reduces to the one presented in Ref.~\cite{Nopoush:2014nc} in the $(0+1)$D case.

A very similar procedure can be used for $\pres^{ij}$, namely
\begin{equation}
 \pres^{ij} = \sqrt{ \det(M^{-1})}\sum_k (M^{-1})^{ik} \left\{ 2 \, \partial_{(M^{-1})^{kj}}\left[ \frac{\ped}{\sqrt{\det(M^{-1})}} \right] \right\},
\end{equation}
and for $\Theta^{ij}$, which gives
\begin{equation}
 \Theta^{ij} = (M^{-1})^{ij} \sqrt{\det(M^{-1})} \; \Thetatreq(\lambda),
\end{equation}
where $\Thetatreq(\lambda)$ is just $\Thetatreq$ with $\lambda$ instead of the equilibrium temperature $T$.

\subsection{Close to equilibrium limit and transport coefficients}
\label{sect:cte}

The close-to-equilibrium formulas~(\ref{shear->eq_2}),~(\ref{zeroth->eq}),~(\ref{bulk->eq}) and~(\ref{UmuUnu->eq}) depend on the the proper energy density $\pedeq$, equilibrium pressure $\preseq$, equilibrium particle number $n_{\rm eq.}$, and the quantities $\Thetatreq$, $g$, $\hzero$, $\htr$, $\phi_6$, $\phi_7$, $\phi_1^{\sf tr.}$, $\phi_3^{\sf tr.}$, $\phi_1^0$ and $\phi_3^0$. I have already shown the explicit formulas for the first three quantities in Eqs.~(\ref{Landau_matching}),~(\ref{neq}) and~(\ref{Peq}). For the other quantities, it is useful to introduce the integrals
\begin{equation}
 I_{n,q} = \alpha^{-1}(q) \int dP \, (p\cdot U)^{n-2q}(p\cdot\Delta\cdot p)^{q}\feq,  \qquad \alpha(q) =(-1)^q (2q+1)!!
\label{Inq}
\end{equation}
In this way, the function $\Thetatreq$ can be written as
\begin{equation}
 \Thetatreq(m,T) = -\alpha(1)\; I_{3,1}(m,T) = 3\; I_{3,1}(m,T).
\label{Thetatreq}
\end{equation}
%
%
%
\textcolor{black}{In order to get the formulas for the other quantities it is necessary to have a second order expansion of the distribution function}

\begin{eqnarray}\nonumber
 f &\simeq& \feq \left\{ 1 + \frac{\lambda -T}{T^2}(p\cdot U) \left[ 1 + \frac{\lambda-T}{T^2}\left( \frac{1}{2}(p\cdot U) -T \right) \right]\right. \\ \nonumber
 && \qquad \qquad  -\frac{1}{2T(p\cdot U)}\left( \frac{}{} \xi_{\mu\nu} - \phi \Delta_{\mu\nu} \right) p^\mu p^\nu\left[ 1 +\frac{\lambda-T}{T^2}\left( \frac{}{} (p\cdot U) -T \right) \right] \\
 && \qquad \qquad \left. +\frac{1}{8T^2(p\cdot U)^3}\left( \frac{}{} \xi_{\mu\nu} - \phi \Delta_{\mu\nu} \right)\left( \frac{}{} \xi_{\rho\sigma} - \phi \Delta_{\rho\sigma} \right) p^\mu p^\nu p^\rho p^\sigma \left[ \frac{}{} (p\cdot U) +T \right]\right\}, \label{f->eq_II}
\end{eqnarray}
\textcolor{black}{Because of symmetry,  all of the moments of $\feq$ must be proportional only to the four velocity $U^\mu$, the projection $\Delta^{\mu\nu}$ and combinations thereof. It can be shown that stress-energy tensor, {\it i.e.} the second moment of the full distribution function $f$, reads}
\begin{eqnarray}\nonumber
 T^{\mu\nu} &\simeq & T^{\mu\nu}_{\rm eq.} \\ \nonumber
 && + \left\{ \frac{\lambda-T}{T^2} \left[ I_{3,0} +\frac{\lambda - T}{T^2} \left( \frac{1}{2} I_{4,0} - T I_{3,0} \right) \right] - \frac{3\phi}{2T} \left[ I_{3,1} +\frac{\lambda -T }{T^2} \left( \frac{}{} I_{4,1} - T I_{3,1} \right)\right] + \right. \\ \nonumber
 && \qquad \left. + \frac{1}{8T^2} \left( \frac{}{} I_{4,2} + T I_{3,2} \right) \left( \frac{}{} 2 \, \xi_{\mu\nu}\xi^{\mu\nu} + 15 \, \phi^2 \right) \right\}U^\mu U^\nu \\ \nonumber
 && +\left\{ - \frac{\lambda-T}{T^2} \left[ I_{3,1} +\frac{\lambda - T}{T^2}\left( \frac{1}{2}I_{4,1} - T I_{3,1} \right) \right] +\frac{5\phi}{2T} \left[ I_{3,2} +\frac{\lambda -T}{T^2}\left( \frac{}{}  I_{4,2} - T  I_{3,2}\right) \right] \right. \\ \nonumber
 && \qquad \left. -\frac{1}{8T^2}\left( \frac{}{} I_{4,3} + T I_{3,3} \right)\left( 35 \, \phi^2+\frac{14}{3} \, \xi_{\mu\nu}\xi^{\mu\nu} \right) \right\}\Delta^{\mu\nu}  \\ 
 && -\frac{1}{T}\left[ I_{3,2} +\frac{\lambda -T}{T^2}\left(\frac{}{} I_{4,2} - T I_{3,2} \right) \right] \xi^{\mu\nu}-\frac{1}{8T^2}\left( I_{4,3} + T I_{3,3} \frac{}{} \right)\left(\frac{}{} 8 \, \xi_\lambda^{\langle\mu}\xi^{\nu\rangle\lambda} - 28 \, \phi \, \xi^{\mu\nu}  \right) . \label{limit_Tmunu}
\end{eqnarray}
\textcolor{black}{after using~(\ref{f->eq_II}), and performing all the integrals and tensor contractions.}
%
%
\textcolor{black}{In order to write simpler formulas it is convenient to use the Landau matching~(\ref{Landau_matching}). If it is plugged it into~(\ref{limit_Tmunu}), it reads}

\begin{eqnarray}\nonumber
  \frac{\lambda-T}{T^2}I_{3,0}  &\simeq& \frac{3\phi}{2T} I_{3,1} +\frac{\lambda -T}{T^2}\left[ \frac{3\phi}{2T}\left( \frac{}{} I_{4,1} -T I_{3,1} \right)-\frac{\lambda-T}{T^2}\left( \frac{1}{2}I_{4,0}- T I_{3,0} \right) \right] \\ \label{Landau->eq}
  && \qquad -\frac{1}{8T^2} \left( \frac{}{} I_{4,2} + T I_{3,2} \right) \left( \frac{}{} 2 \, \xi_{\mu\nu}\xi^{\mu\nu} + 15 \, \phi^2 \right).
\end{eqnarray}
\textcolor{black}{The last approximation is the Landau matching, up to second order in deviations from equilibrium. The first order is }
\begin{equation}
 \frac{\lambda-T}{T^2}  \simeq \frac{3\phi}{2T} \, c_s^2,
\label{Landau->eq_I}
\end{equation}
\textcolor{black}{where $c_s^2 = I_{3,1}/I_{3,0} = \partial \preseq/\partial \pedeq$ is the square of the speed of sound.}

\textcolor{black}{In the right hand side of Eq.~(\ref{Landau->eq}) the difference $\lambda-T$ appears only on second order terms. Therefore, plugging the first order approximation~(\ref{Landau->eq_I}) into the right hand side of the second order approximation~(\ref{Landau->eq}), the Landau Matching reads}

\begin{eqnarray}\label{Landau->eq_II}
  \frac{\lambda-T}{T^2}  &\simeq& \frac{3\phi}{2T}\, c_s^2 +\frac{9 \phi^2}{4T^2}\, c_s^2 \left( \frac{I_{4,1}}{I_{3,0}} - \frac{1}{2}\, c_s^2\,  \frac{I_{4,0}}{I_{3,0}} \right) -\frac{1}{8T^2} \left( \frac{I_{4,2}}{I_{3,0}} + T \frac{I_{3,2}}{I_{3,0}} \right) \left( \frac{}{} 2 \, \xi_{\mu\nu}\xi^{\mu\nu} + 15 \, \phi^2 \right).
\end{eqnarray}
\textcolor{black}{Plugging the Landau matching~(\ref{Landau->eq_II}) in Eq.~(\ref{limit_Tmunu}) the stress-energy-momentum tensor reads}

\begin{equation}\label{Tmunu_II}
 T^{\mu\nu}\simeq T^{\mu\nu}_{\rm eq} + \varPi_\phi \; \phi \, \Delta^{\mu\nu} + \varPi_{\phi\phi} \; \phi^2 \, \Delta^{\mu\nu} + \varPi_{\xi\xi} \; \xi_{\rho\sigma}\xi^{\rho\sigma} \, \Delta^{\mu\nu} - \varpi_{\xi} \; \xi^{\mu\nu} - \varpi_{\phi\xi} \, \phi\; \xi^{\mu\nu} - \varpi_{\xi\xi} \; \xi_\lambda^{\langle\mu}\pi^{\nu\rangle\lambda}.
\end{equation}
\textcolor{black}{Recognizing then}

\begin{equation}
 \Pi \simeq -\varPi_\phi \; \phi  - \varPi_{\phi\phi} \; \phi^2 - \varPi_{\xi\xi} \; \xi_{\rho\sigma}\xi^{\rho\sigma},  \qquad \qquad \pi^{\mu\nu}\simeq- \varpi_{\xi} \; \xi^{\mu\nu} - \varpi_{\phi\xi} \, \phi\; \xi^{\mu\nu} - \varpi_{\xi\xi} \; \xi_\lambda^{\langle\mu}\xi^{\nu\rangle\lambda},
\end{equation}
\textcolor{black}{where}

\begin{equation}\label{Pi}
 \begin{cases}
  \varPi_\phi = \frac{1}{2T}\left[ 5 I_{3,2} -3 \, c_s^2 \, I_{3,1} \right], \\
  \varPi_{\phi\phi} = \frac{1}{8T^2}\left[ 15\, c_s^2 \left( \frac{}{} 3 I_{4,2} - T I_{3,2} \right) - 18\, c_s^4 \left( \frac{3}{2}I_{4,1} -\frac{1}{2}\, c_s^2 \, I_{4,0} - T I_{3,1} \right) -35 \left(  \frac{}{}  I_{4,3} + T I_{3,3}\right) \right] ,\\
  \varPi_{\xi\xi} = \frac{1}{12T^2}\left[ 3\, c_s^2 \left( I_{4,2} + T I_{3,2} \right) - 7 \left( I_{4,3} + T I_{3,3} \right) \right] ,
 \end{cases}
\end{equation}
\textcolor{black}{and}

\begin{equation}\label{pi}
 \begin{cases}
  \varpi_\xi = \frac{1}{T} I_{3,2}, \\
  \varpi_{\phi\xi} = \frac{1}{2T^2}\left[ 3\, c_s^2 \left( \frac{}{} I_{4,2} - T I_{3,2} \right) - 7 \left( \frac{}{} I_{4,3} + T I_{3,3} \right) \right],  \\
  \varpi_{\xi\xi} = \frac{1}{T^2} \left( \frac{}{} I_{4,3} + T I_{3,3} \right).
 \end{cases}
\end{equation}
\textcolor{black}{Naturally, the first order functions $\varPi_\phi$ and $\varpi_\xi$ could have been obtained just using the linear expansion of the distribution function~(\ref{a-Hydro_distribution->eq}).}
%
%
\textcolor{black}{All of the functions $\varPi_\phi$, $\varPi_{\phi\phi}$ and $\varPi_{\xi\xi}$ are vanishing in the massless limit as expected.}

\textcolor{black}{Eq.~(\ref{q->eq_II}) stem from the expansion of $\Theta^{\mu\nu}$ and $n$, making use of  Eq.~(\ref{f->eq_II}), the Landau matching~(\ref{Landau->eq_II}),~(\ref{Pi}) and~(\ref{pi}). Where}

%
%
%
%
%
\begin{equation}
 \htr = 3 \,\frac{5 \, I_{4,2} - 3\, c_s^2 \, I_{4,1}  }{ 5I_{3,2} - 3 \, c_s^2 \,I_{3,1} } , \qquad g= \frac{ I_{4,2}}{ I_{3,2}}, \qquad \hzero= 3 \, \frac{ \pedeq \, c_s^2- \preseq}{5I_{3,2} - 3 \, c_s^2 \, I_{3,1} }.
\end{equation}
\begin{equation}
 \begin{cases}
  \phi_6 = \left[ 3\, c_s^2 \left( \frac{}{} I_{5,2} - TI_{4,2}\right)-7\left( \frac{}{} I_{5,3} +T I_{4,3}\right) -2 \, g \, T^2 \, \varpi_{\phi\xi} \right]\left( \frac{}{} 2 \, T^2 \, \varPi_\phi \,  \varpi_\xi \,  g \right)^{-1},\\ \\ 
  \phi_7 = \left[ \frac{}{} I_{5,3} + T I_{4,3} - g\, T^2 \, \varpi_{\xi\xi} \right] ( T^2 \, \varpi_{\xi}^2 \, g )^{-1},
 \end{cases}
\end{equation}
\begin{equation}
 \begin{cases}
  \phi_1^{\sf tr.} =3 \left[ 15\, c_s^2 \left( \frac{I_{4,1}}{I_{3,1}} I_{4,2} + T \frac{I_{4,1}}{I_{3,1}}I_{3,2} + 2I_{5,2} - 2T I_{4,2} \right) -18\, c_s^4 \left( \frac{I_{4,1}}{I_{3,1}}I_{4,1} - \frac{1}{2}\frac{I_{4,1}}{I_{3,0}} I_{4,0} + \frac{1}{2}I_{5,1} - T I_{4,1} \right)\right. \\
  \qquad \qquad \quad \left. -35\left(\frac{}{} I_{5,3} + T I_{4,3} \right) - \frac{8\, T^2 }{3}\, \htr \, \varPi_{\phi\phi}\right] (8 \, T^2 \, \varPi_\phi^2 \, \htr)^{-1} \\ \\
  \phi_3^{\sf tr.} =  \left[ 3\frac{I_{4,1}}{I_{3,0}}\left( I_{4,2} + T I_{3,2} \right) -7 \left( I_{5,3} +T I_{4,3} \right) - 4\, T^2 \, \htr \varPi_{\xi\xi}\right] ( 4\, T^2 \, \varpi_\xi^2 \, \htr )^{-1},
 \end{cases}
\end{equation}
\begin{equation}
 \begin{cases}
  \phi_1^{0} =  \left[ 18\, c_s^4 \left( \frac{I_{4,1}}{I_{3,1}} \, \pedeq  - \frac{1}{2}\frac{I_{4,0}}{I_{3,0}} \, \pedeq + T\frac{\preseq}{c_s^2}  -\frac{1}{2}I_{3,0} - T \, \pedeq \right) -15\left( \frac{I_{4,2}}{I_{3,0}} \, \pedeq + T \, \frac{I_{3,2}}{I_{3,0}} \, \pedeq - I_{3,2} -T I_{2,2}  \right) \right. \\
 \qquad \qquad \qquad \qquad \left.  -8 \, T^2 \, \hzero \, \varPi_{\phi\phi}\right] (8 \, T^2 \, \varPi_{\phi}^2 \, \hzero)^{-1}\\ \\
  \phi_3^{0} = \left( I_{3,2} + T I_{2,2} - \frac{I_{4,2}}{I_{3,0}} \, \pedeq -T \frac{I_{3,2}}{I_{3,0}} \, \pedeq - 4 \, T^2 \, \hzero \, \varPi_{\xi\xi}\right) (4 \, T^2 \, \varpi_{\xi}^2 \, \hzero)^{-1}.
 \end{cases}
\end{equation}
%

%
%
%
%
%

Finally,there are the analytic formulas for the integrals $I_{n,q}$ which are necessary for the transport coefficients\textcolor{black}{. Reminding that $I_{2,0} = {\pedeq}$ and $I_{2,1}=\preseq$}
\begin{eqnarray}
 && I_{2,2} = (4\pi \tildeN) \frac{T^4 \mheq^2}{30}\left\{ \left( \frac{}{} 6 - \mheq^2 \right) K_2(\mheq) + \mheq^2 \left[ \frac{}{} 3 K_0(\mheq) -2 \mheq \left( \frac{}{} K_1(\mheq) - K_{i1}\right)\right]\right\} \\     
 && I_{3,0} = (4\pi \tildeN) \,T^5 \, \mheq \left[ \mheq \left( \frac{}{} \mheq^2+12 \right)K_0(\mheq) + \left( \frac{}{} 5 \, \mheq^2 +24 \right)K_1(\mheq)\right], \\
 && I_{3,1} = (4\pi \tildeN) \,T^5 \, \mheq^3 \, K_3(\mheq), \\
 && I_{3,2} = \frac{(4\pi \tildeN) \,T^5 \, \mheq^3 }{15} \left\{ \mheq^2\left[ K_1(\mheq) -K_{i1}(\mheq) \frac{}{}\right] - \mheq \, K_2(\mheq) +3\, K_3(\mheq) \right\}, \\
 && I_{3,3} = \frac{(4\pi \tildeN) \,T^5 \, \mheq^5 }{105} \frac{1}{2}\left\{ \frac{15}{\mheq}K_4(\mheq) \right. \\
 &&\qquad \left.-\left( \frac{84}{\mheq^2} +3 \right)K_3(\mheq)- \left( \frac{}{} 7-\mheq^2 \right)\left[ \frac{1}{\mheq}K_2(\mheq) -K_1(\mheq) +K_{i1}(\mheq)  \right] \right\}, \\
 && I_{4,1} = (4\pi \tilde N) \,T^6 \, \mheq \left[ \mheq\left(\frac{}{} \mheq^2 + 20\right) K_0(\mheq) + \left( \frac{}{} 7 \mheq^2 +40 \right) K_1(\mheq)\right], \\
 && I_{4,2} = (4\pi \tildeN) \,T^6 \, \mheq^3 \, K_3(\mheq), \\
 && I_{5,1} = (4\pi \tildeN) \,T^7 \, \mheq^3 \left[ \left( \frac{}{} 30 + \mheq^2 \right)K_3(\mheq) + 5 \, \mheq K_{2}(\mheq) \right], \\
 && I_{5,2} = (4\pi \tildeN) \,T^7 \, \mheq^3 \left[ \frac{}{} 6K_3(\mheq) +\mheq K_2(\mheq) \right],
\end{eqnarray}
where $K_n$'s are the modified Bessel functions and
\begin{equation}
 K_{i1}(x) = \int_0^\infty \!\!\! d\eta \frac{e^{-x\cosh\eta}}{\cosh\eta} = \frac{\pi}{2}\left[ \frac{}{} 1 - x \, K_0(x)\, L_{-1}(x) - x \, K_1(x)\, L_{0}(x)  \right]
\end{equation}
with $ L_{n}(x)$ being a modified Struve function.

\textcolor{black}{The remaining integrals can be written as a combination of the previous ones}

\begin{eqnarray}
 && I_{4,0} = 3I_{4,1} + T^2 \mheq^2I_{2,0}, \\
 && I_{4,3} = \frac{1}{7}\left( \frac{}{} I_{4,2}- T^2\mheq^2 I_{2,2} \right), \\
 && I_{5,3} = \frac{1}{7}\left( \frac{}{} I_{5,2}- T^2\mheq^2 I_{3,2} \right).
 \end{eqnarray}

\end{document}